\documentstyle[aps,prl,epsfig,floats]{revtex}
\def\NdCe{Nd$_{1.85}$Ce$_{0.15}$CuO$_4$\,}
\def\cm-1{cm$^{-1}$}
\def\Tc{T$_{c}$\,}
\begin{document}
\draft

\title{Low-energy excitations around $(\frac{\pi}{2}, 
\frac{\pi}{2})$ points in the pseudogap phase of \NdCe }

\author{
A.~Koitzsch$^{1,\dag}$, G.~Blumberg$^{1,\S}$, A.~Gozar$^{1}$,
B.S.~Dennis$^{1}$, P.~Fournier$^{2,\P}$, and R.L.~Greene$^{2}$
}
\address{
$^{1}$Bell Laboratories, Lucent Technologies, Murray Hill, New Jersey
07974 \\
$^{2}$Center for Superconductivity Research and Department of Physics,
University of Maryland, College Park, MD 20742 \\
}

\twocolumn[\hsize\textwidth\columnwidth\hsize\csname
@twocolumnfalse\endcsname
\date{\today}
\maketitle
\widetext

\begin{abstract}

      Polarized electronic Raman scattering from the \NdCe superconductor 
      at optimal electron doping ($T_{c}=22$~K) reveals the 
      formation of an anisotropic pseudogap below a characteristic 
      temperature $T^{*} \approx 220$~K and energy $E^{*}_{g} = 850$~\cm-1.
      Below $T^{*}$ a pronounced suppression of the incoherent spectral 
      weight below $E^{*}_{g}$ in the nodal directions 
      of \textbf{k}-space is observed. 
      This is concomitant with the emergence of long-lived excitations 
      in the vicinity of the $(\pm\pi/2, \pm\pi/2)$ points that do not 
      contribute to the optical conductivity. 

\end{abstract}

]

\narrowtext

The origin of the pseudogap (PG) in cuprate superconductors is
believed to lie at the heart of a general understanding of superconductivity
in this material class.
A large body of experimental results have been collected for hole
doped cuprates  and many theoretical approaches have been 
developed \cite{Timusk}.
A consensus on the underlaying  mechanism is still far from
being reached.
Recently studies of electron doped cuprates have shifted under focus, 
first, with the reexamination of the symmetry of the order
parameter \cite{Tsuei,Prozorov,Sato,Armitage1,Blumberg,Biswas2,Skinta}
whereby a $d_{x^2-y^2}$ symmetry was established for under- and
optimal doping, and second, with the investigation of the PG
\cite{Homes,Singley,Armitage2,Onose,Tohyama}.
It was also found that electron doped cuprates violate
the Wiedemann-Franz law, indicating the breakdown of Fermi liquid
theory \cite{Hill}.
These materials offer the unique possibility of comparison with their
hole doped counterparts which can serve as a touchstone for theory.
Furthermore, electron doped materials like \NdCe (NCCO) show their own
peculiarities.
The doping range where the superconducting (SC) transition occurs is
much narrower than for hole doped cuprates and the maximum achievable
SC transition temperatures (\Tc) are much lower.
Also, the normal state in-plane-resistivity shows a tendency to a
quadratic
temperature dependence \cite{Tsuei89} in contrast to the optimally
hole doped cuprates where the dependence is linear
\cite{Gurvich}.

While PG phenomena for hole doped cuprates was investigated by a large
number of experimental probes the 
evidence for a PG in electron doped cuprates is still limited. 
A recent optics study has found a large PG for underdoped NCCO 
\cite{Onose}. 
For optimally doped NCCO it has been shown that the scattering 
rate extracted from optical conductivity, $\sigma(\omega)$, is suppressed 
below certain energies and temperatures \cite{Homes,Singley}.  
The ARPES experiment showed regions of suppressed intensity at the
Fermi surface (FS) with a broad peak at about 100~meV ($\approx
800$~\cm-1) at low temperatures \cite{Armitage2}.
These regions are shifted closer to the nodal directions with
respect to the hole doped cuprates.
On the other hand, $c$-axis tunneling studies did not indicate PG 
suppression above \Tc \cite{Kleefisch,Biswas}. 

Here we report a Raman study of NCCO in the PG phase.
In contrast to the observations for hole doped cuprates, for NCCO, 
a substantial suppression of the Raman response below an onset 
temperature $T^{*} \approx 220$~K and a threshold energy $E^{*}_{g} = 
850$~\cm-1 was observed for the symmetry that probes excitations in the 
proximity of the nodal directions of \textbf{k}-space.
Furthermore, we found novel long lived low energy excitations that 
reside in the vicinity of $(\pm\pi/2, \pm\pi/2)$ regions in the PG 
phase but do not show an observable contribution to the optical 
conductivity.       
The excitations in the proximity of the Brillouin zone (BZ) boundaries 
observed by Raman scale to the optical conductivity data. 

The Raman experiments were performed from a natural $ab$ surface of 
plate-like single crystals grown as described in Ref. \cite{Peng}.
After growth, the crystals were annealed in an oxygen-reduced atmosphere
to induce the doping level for an optimal \Tc.
The SC transition measured by SQUID was about 22~K with a width of 
about 2~K.
The sample was mounted in an optical continuous helium flow cryostat.
Spectra were measured in a backscattering geometry using the 6471~\AA \
excitation of a Kr$^{+}$ laser.
An incident laser power less than 10~mW was focused to a 50~$\mu$m spot
on the sample surface.
The referred temperatures were corrected for estimated laser heating.
The spectra were analyzed by a custom triple grating spectrometer and
were corrected for the instrumental response.

The presented data were taken in ($xy$) and ($x^{\prime}y^{\prime}$)
scattering geometries, with $x = [100]$, $y = [010]$,
$x^{\prime} = [110]$, and $y^{\prime} = [\bar{1}10]$.
For tetragonal D$_{4h}$ symmetry the ($xy$) and
($x^{\prime}y^{\prime}$) geometries correspond to spectra of
B$_{2g}$ + A$_{2g}$ and B$_{1g}$ + A$_{2g}$ representations.
Measurements in an external magnetic field have been performed with
circularly polarized light where a right-left (RL) scattering geometry
corresponds to spectra of B$_{1g}$ + B$_{2g}$ representation.
In addition, by using right-right (RR) geometry, we checked
the intensity of the A$_{2g}$ component and found it to be negligibly 
weak.

The electronic Raman response function for a given scattering geometry
is proportional to the sum over the density of states at the FS 
weighted by a momentum \textbf{k} dependent Raman form factor.
By choosing the scattering geometry one can selectively probe 
different regions of the FS.
In the insets of Fig.~1 we sketch the first BZ for NCCO as suggested 
by ARPES data \cite{Sato,Armitage1,Armitage2}. 
The shaded area corresponds to occupied electron states.
For the B$_{2g}$ channel the Raman form factor vanishes along
$(0, 0) \rightarrow (\pi, 0)$ and the equivalent
lines of \textbf{k}-space. 
The Raman response projects out excitations in the vicinity of  
$(\pm\pi/2, \pm\pi/2)$ points of the BZ.
In contrast, the B$_{1g}$ intensity vanishes along the diagonals and
the intensity is collected from the regions of the BZ distant from the
diagonals.
The $45^{\circ}$ rotated square indicates the antiferromagnetic 
Brillouin zone (AF BZ).
Intersections of the FS and the AF BZ are marked in the upper left
quadrant with black circles.
These points are connected by the AF unit vector $(\pi, \pi)$.

\begin{figure}[t]
\begin{center}
\epsfig{file=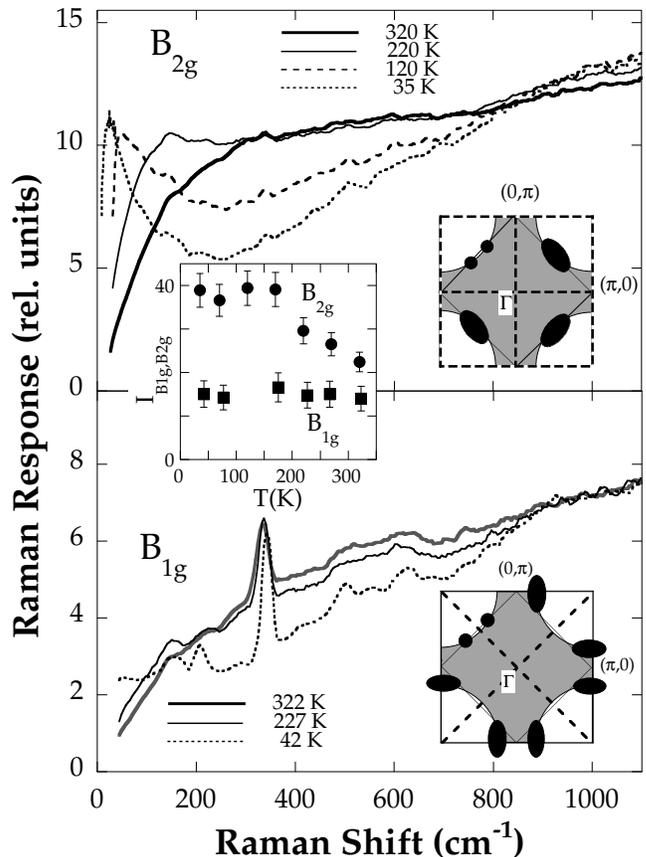, width=8.5cm}
\caption[]{
Raman response function for $B_{2g}$ and $B_{1g}$ symmetry channels 
(1.9~eV excitation) at temperatures between 35 and 320~K.
Sketches for the Fermi surface are based on ARPES data 
\cite{Sato,Armitage1,Armitage2}.
Regions where the Raman form-factor is maximal 
are denoted by black ellipses; the nodal regions are shown by dashed lines.
The antiferromagnetic Brillouin zone is indicated as a square rotated
by $45^{\circ}$.
AF fluctuations enhance interactions between fermions around the 
\emph{hot spots} (filled circles).
The sharp modes for $B_{1g}$ are phonons \cite{Heyen}. 
The inset shows $I_{\mu} = \int_{0}^{850}\chi''_{\mu}/\omega\,d\omega$ 
for $\mu = B_{1g}$ (squares) and $B_{2g}$ (circles).  
}
\label{Chi}
\end{center}
\end{figure}

In Fig.~1 we show the Raman response as a function of temperature for
the B$_{1g}$ and B$_{2g}$ channels.
For B$_{2g}$ symmetry two distinct features develop in the spectra
with cooling:

(i) A "knee" like intensity loss of the Raman response below
850~\cm-1.
This suppression of spectral weight starts at $T^{*}_{B_{2g}} \approx 
220$~K.  
It deepens with cooling and decreasing energy yielding an almost
linear
slope of the spectra between 300 and 850~\cm-1.
We identify this suppression as the opening of a PG with a
temperature independent energy scale.

(ii) An overdamped peak is seen at about 350~\cm-1 at room
temperature.
This peak sharpens with cooling and shifts to
lower frequencies from 150~\cm-1 at 220~K to 25~\cm-1 at 35~K.
We refer to this feature as a quasi-elastic scattering peak (QEP).

The opening of the PG with a similar energy scale is also present but
not as pronounced in the B$_{1g}$ channel.
The QEP however is much weaker and at higher energies. 

In contrast, for hole doped cuprates the PG suppression is most 
pronounced for the B$_{1g}$ channel
\cite{Slakey90,Blumberg97,Chen,Quilty,Naeini,B"ckstr–m}.
For hole underdoped cuprates a destruction of the FS near its
intersection with AF BZ boundary has been observed 
\cite{Marshall,Norman}.
Therefore Raman scattering for the B$_{1g}$ symmetry is suppressed
leaving only small PG effects for the B$_{2g}$ channel
\cite{Blumberg97,Nemetschek,Opel}.

A conjecture that the PG opens up at the intersections of the AF BZ 
and FS, the \emph{hot spots} \cite {Hlubina,Chubukov,Schmalian,Ioffe}, 
reconciles the observed differences for the electron and hole doped 
materials. 
Strong antiferromagnetic interactions suppress the spectral weight at 
the chemical potential in the \emph{hot spots} leading to a 
"destruction" of the FS in this region of \textbf{k}-space and opening 
of the PG. 
In hole doped cuprates \emph{hot spots} are located closer to the 
($\pi$, 0) points while for electron doping the FS shrinks shifting
\emph{hot spots} closer to the nodal directions
\cite{Blumberg,Armitage2,Marshall}.
Hence, for the hole underdoped systems the PG is more
pronounced in B$_{1g}$ while for electron doped systems in
the B$_{2g}$ scattering channel.

Cooling the sample below T$_c = 22$~K leads to opening of the
SC gap and formation of a pair breaking 2$\Delta$
peak \cite{Blumberg}.
The QEP is suppressed in this regime.
An applied magnetic field up to 9~T did not influence either
the PG or the QEP above \Tc .
In contrast, below \Tc an external magnetic field of this magnitude
strongly suppresses SC correlations \cite{BluPre} and reveals the 
underlying QEP (See Fig.~2).

\begin{figure}[h]
\begin{center}
\epsfig{file=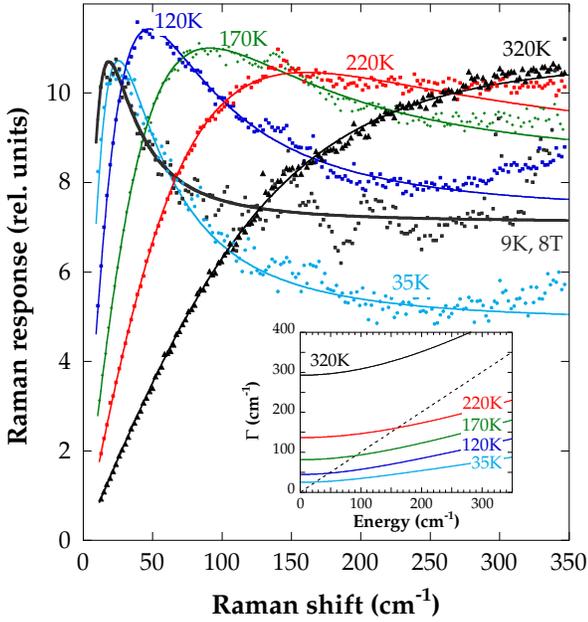, width=8.0cm}
\vspace{3mm}
\end{center}
\caption[]
{Low frequency Raman response for the $B_{2g}$ channel
between 35 and 320~K.
The quasi-elastic peak develops with cooling.
The 9~K spectra was measured in an external magnetic field applied
parallel to the $c$-axis in RL-polarization ($B_{1g}$ +$B_{2g}$
channel). 
The solid lines are fits to the extended Drude model (\ref{QEP}).
Inset: Frequency dependent scattering rate $\Gamma^{B_{2g}}(\omega,
T)$ resulting from the fits to the Drude model (\ref{QEP})
for temperatures between 35 and 320~K.
The dashed $\Gamma = \omega$ line represents 
a crossover from an overdamped regime to longer lived excitations.
}
\label{Int}
\end{figure}

In Fig.~2 we show the low frequency part of the Raman response in the
B$_{2g}$ channel.
The QEP develops with decreasing temperature.
It is also present at
temperatures below \Tc if superconductivity is suppressed by a 
magnetic field \cite{comment2}.
We fit the low energy B$_{2g}$ response by an extended Drude model:
\begin{equation}
\chi''(\omega, T) \propto N^{B_{2g}}_{F} \frac{\omega 
\Gamma^{B_{2g}}(\omega,
T)}{\omega^{2} + \Gamma^{B_{2g}}(\omega, T)^{2}},
\label{QEP}
\end{equation}
where $N^{B_{2g}}_{F}$ is the density of states at the Fermi level 
weighted by a symmetry dependent Raman form factor and
$\Gamma^{B_{2g}}(\omega, T)$
is a frequency and energy dependent scattering rate.
The scattering rate $\Gamma^{B_{2g}}(\omega, T)$ for different
temperatures extracted from the extended Drude fits is shown in the
inset of Fig.~2.
$\Gamma^{B_{2g}}$ is about 300~\cm-1 at room temperature and decreases 
rapidly with cooling.
The small variation with frequency suggests that a simple Drude model is 
a good description of the data.
However, qualitative differences appear when the
momentum dependence of the scattering rate is analyzed.

\begin{figure}[h]
\begin{center}
\epsfig{file=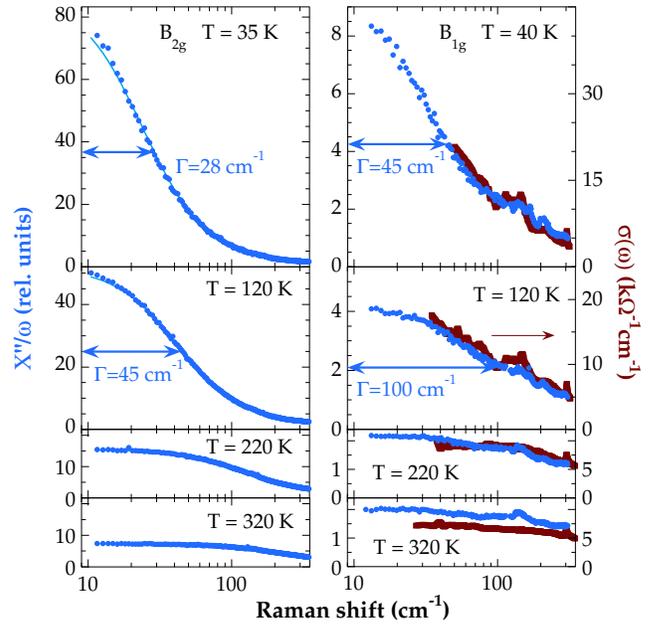, width=8.5cm}
\vspace{3mm}
\end{center}
\caption[]
{$\chi''_{\mu}/\omega$ response for $\mu = B_{2g}$ and $B_{1g}$ 
channels (dots in left and right panels) between 35~K and 320~K.
The optical conductivity (thick dark solid line) is shown on the right 
scale \cite{Homes} and is consistent with $\chi''_{B_{1g}}/\omega$.
The thin solid lines on the left represent the fit to the extended Drude 
model (1).
Phonons have been removed from the $B_{1g}$ spectra.
}
\label{rho}
\end{figure}

In Fig.~3 we compare optical conductivity and $\chi''_{\mu}/\omega$ 
for $\mu = B_{2g}$ and $B_{1g}$ channels between 35 and 300~K.
In the lowest order approximation $\chi''_{\mu}(\omega)/\omega$ is 
proportional to optical conductivity weighted by a geometrical Raman 
form factor \cite{Sriram}.
We notice that the Drude response narrows with cooling for 
$\sigma(\omega)$ as well as for $\chi''_{\mu}/\omega$ in both $\mu = 
B_{2g}$ and $B_{1g}$ channels. 
For B$_{1g}$ the width of the peak decreases from above 400~\cm-1 around
room temperature to 45~\cm-1 above the superconducting transition,
retaining however a value about twice that of the B$_{2g}$ channel at
corresponding temperatures.
This difference implies a momentum dependence of the scattering rate:  
a longer lifetime for excitations in the vicinity of $(\pm\pi/2, 
\pm\pi/2)$ points and more incoherence for excitations around the BZ 
boundaries.   
Our observation of increased coherence in the B$_{2g}$ channel 
is consistent with ARPES results, where the quasi-particle 
peak linewidth at the Fermi energy decreases in going from the BZ 
boundary to $(\pi/2, \pi/2)$ point \cite{Armitage2}.

The momentum dependence of the scattering rate is  consistent with the 
high frequency Raman data (Fig.~1).
In the inset of Fig.~1 we present 
$I_{\mu} = \int_{0}^{850}\chi''_{\mu}/\omega\,d\omega$ 
($\mu = B_{1g}, B_{2g}$) as a function of temperature \cite{sumrule}.  
While $I_{B_{1g}}$ is a constant $I_{B_{2g}}$ increases with cooling 
until 170~K indicating a shift of spectral weight from regions beyond 
850~\cm-1 to the lower frequencies \cite{sumrule}.
The pronounced suppression of the $B_{2g}$ Raman continuum intensity is
another indication of increased coherence at low energies for 
excitations in the vicinity of $(\pm\pi/2, \pm\pi/2)$ points.

On the right side of Fig.~3 we also plot optical conductivity from 
Ref.~\cite{Homes}. 
Surprisingly, $\sigma(\omega)$ exhibits almost perfect scaling to 
$\chi''_{B_{1g}}(\omega)/\omega$ Raman data: 
the scaling factor for the double vertical axes was determined for 
T~=~120~K data and then used for all temperatures.
The fact that the B$_{1g}$ response bears out the pseudoidentity
Re~$\sigma \propto \chi''_{B_{1g}}(\omega)/\omega$ \cite{Sriram} suggests 
that quasi-particle contributions to the low frequency and $dc$ 
conductivities are dominated by parts of the Fermi
surface away from the $(0, 0) \rightarrow (\pi, \pi)$ diagonal.
The more coherent excitations in the vicinity of $(\pm\pi/2, \pm\pi/2)$ 
points do not contribute significantly to the optical response. 
Nevertheless the latter excitations dominate the low frequency B$_{2g}$ 
Raman response in the PG state.
Although further studies are necessary, we suggest that these 
chargeless excitations may be responsible for the excessive heat transport 
that leads to the Wiedemann-Franz law violation \cite{Hill}.

In summary, we study charge carrier relaxation dynamics in the PG phase 
of NCCO.  
We observe suppression of spectral weight below 850~\cm-1 for the 
B$_{2g}$ Raman response and identify it as an anisotropic
PG in the vicinity of $(\pm\pi/2, \pm\pi/2)$ points of the BZ.
We propose that the PG originates from enhanced AF interactions in
\emph{hot spot} regions, which for electron doped cuprates, in
contrast to hole doped, are located closer
to $(\pm\pi/2, \pm\pi/2)$ points than to the BZ boundary.
For the Raman response in the B$_{2g}$ channel we observe a narrow 
Drude-like QEP in the PG phase.
This QEP reveals the emergence of novel long lived excitations in the 
vicinity of the $(\pm\pi/2, \pm\pi/2)$ points that do 
not contribute to optical conductivity. 
In contrast, the excitations in the $B_{1g}$ Raman response were found 
to be in agreement with optical conductivity data. 

AK acknowledges partial support by the Studienstiftung des Deutschen
Volkes. RLG acknowledges support by NSF grant DMR-0102350.
PF acknowledges the support from CIAR, NSERC and the Foundation
FORCE (Sherbrooke).
We thank C. Homes for providing the optical conductivity data. AK  
acknowledges discussions with S.V.~Borisenko and T.~Pichler.


\begin{references}

\vspace{-15mm}

\bibitem[\S]{byline}  To whom correspondence should be addressed.
E-mail: girsh@bell-labs.com

\bibitem[\dag]{Koitzsch} Permanent address: Institute of Solid State and
Materials Research Dresden, P.O. Box 270016, D-01171 Dresden, Germany.

\bibitem[\P]{Fournier} Permanent address: Centre de recherche sur les
propri\'et\'es \'electroniques de mat\'eriaux avanc\'es and
D\'epartement
de Physique, Universit\'e de Sherbrooke, Sherbrooke, Qu\'ebec,
CANADA, J1K 2R1.

\bibitem{Timusk} T. Timusk and B.Statt,
Rep. Prog. Phys.\textbf{62}, 61 (1999).

\bibitem{Tsuei} C.C. Tsuei and J.R. Kirtley,
Phys. Rev. Lett. \textbf{85}, 182 (2000).

\bibitem{Prozorov} R. Prozorov \textit{et al.},
Phys. Rev. Lett. \textbf{85}, 3700 (2000);
J.D. Kokales \textit{et al.},
Phys. Rev. Lett. \textbf{85}, 3696 (2000).

\bibitem{Sato} T. Sato \textit{et al.},
Science \textbf{291}, 1517 (2001).

\bibitem{Armitage1} N.P. Armitage \textit{et al.},
Phys. Rev. Lett. \textbf{86}, 1126 (2001).

\bibitem{Blumberg} G. Blumberg \textit{et al.},
Phys. Rev. Lett. \textbf{88}, 107002 (2002).

\bibitem{Biswas2} A. Biswas \textit{et al.},
Phys. Rev. Lett. \textbf{88}, 207004 (2002).

\bibitem{Skinta} J.A. Skinta \textit{et al.},
Phys. Rev. Lett. \textbf{88}, 207005 (2002).

\bibitem{Homes} C.C. Homes \textit{et al.},
Phys. Rev. B \textbf{56}, 5525 (1997).

\bibitem{Singley} E.J. Singley \textit{et al.},
Phys. Rev. B \textbf{64}, 224503 (2001).

\bibitem{Armitage2} N.P. Armitage \textit{et al.},
Phys. Rev. Lett. \textbf{87}, 147003 (2001).  

\bibitem{Onose} Y. Onose \textit{et al.},
Phys. Rev. Lett. \textbf{82}, 5120 (1999); 
Phys. Rev. Lett. \textbf{87}, 217001 (2001).

\bibitem{Tohyama} T. Tohyama and S. Maekawa, cond-mat/0106311 (2001).

\bibitem{Hill} R.W. Hill \textit{et al.},
Nature \textbf{414}, 711 (2001).

\bibitem{Tsuei89} C.C. Tsuei \textit{et al.},
Physica C \textbf{161}, 415 (1989).

\bibitem{Gurvich} M. Gurvich \textit{et al.},
Phys. Rev. Lett. \textbf{59}, 1337 (1987).

\bibitem{Kleefisch} S. Kleefisch \textit{et al.},
Phys. Rev. B. \textbf{63}, 100507 (2001).

\bibitem{Biswas} A. Biswas \textit{et al.},
Phys. Rev. B. \textbf{64}, 104519 (2001).

\bibitem{Peng} J.L. Peng \textit{et al.},
Physica C \textbf{177}, 79 (1991).

\bibitem{Heyen} E.T. Heyen \textit{et al.},
Phys. Rev. B \textbf{43}, 2857 (1991).

\bibitem{Slakey90} F. Slakey \textit{et al.},
Phys. Rev. B \textbf{42}, 2643 (1990).

\bibitem{Blumberg97} G. Blumberg \textit{et al.},
Science \textbf{278}, 1427 (1997).

\bibitem{Chen} X.K. Chen \textit{et al.},
Phys. Rev. B \textbf{56}, R513 (1997).

\bibitem{Quilty} J.W. Quilty \textit{et al.},
Phys. Rev. B \textbf{57}, R11097 (1998).

\bibitem{Naeini} J.G. Naeini \textit{et al.},
Phys. Rev. B \textbf{59}, 9642 (1999).

\bibitem{B"ckstr–m} J. B\"ackstr\"om \textit{et al.},
Phys. Rev. B \textbf{61}, 7049 (2000).

\bibitem{Marshall} D.S. Marshall \textit{et al.},
Phys. Rev. Lett. \textbf{76}, 4841 (1996).

\bibitem{Norman} M.R. Norman \textit{et al.},
Nature (London) \textbf{392}, 157 (1998).

\bibitem{Nemetschek} R. Nemetschek \textit{et al.},
Phys. Rev. Lett. \textbf{78}, 4837 (1997).

\bibitem{Opel} M. Opel \textit{et al.},
Phys. Rev. B \textbf{61}, 9752 (2000).

\bibitem{Hlubina} R. Hlubina and T.M. Rice,
Phys. Rev. B \textbf{51}, 9253 (1995).

\bibitem{Chubukov} A. Chubukov,
Europhys. Lett. \textbf{44}, 655 (1997).

\bibitem{Schmalian} J. Schmalian \textit{et al.},
Phys. Rev. Lett. \textbf{80}, 3839 (1998).

\bibitem{Ioffe} L.B. Ioffe and A.J. Millis,
Phys. Rev. B \textbf{58}, 11631 (1998).

\bibitem{BluPre} G. Blumberg \textit{et al.},
preprint

\bibitem{comment2} 
The different background for the 9~K/8~T spectrum is due to RL 
polarization (B$_{2g}$ + B$_{1g}$) used for scattering in magnetic 
field.  

\bibitem{Sriram}
B. S. Shastry, and B. I. Shraiman, Phys. Rev. Lett. {\bf 65}, 1068
(1990).

\bibitem{sumrule}
In analogy to the sum rules for optical conductivity 
$I^{\inf}_{\mu} = \int_{0}^{\inf}\chi''_{\mu}/\omega\,d\omega$ 
is expected to be a temperature independent constant. 

\end{references}
\end{document}